\documentclass[aps,twocolumn,prd,preprintnumbers,superscriptaddress,bibnotes,floatfix,longbibliography]{revtex4-1}

\pdfoutput=1
\usepackage{amsmath}
\usepackage{amsfonts}
\usepackage{amssymb}
\usepackage{mathrsfs}
\usepackage{color}
\usepackage{bm}
\usepackage{graphicx}
\usepackage{blindtext}
\usepackage{wasysym}
\usepackage{mwe}
\usepackage{hyperref}
\usepackage[normalem]{ulem}
\usepackage{multirow}
\usepackage{array}

\newcommand{\ie}{{i.e.}}

\newcommand{\eg}{{e.g.}}

\newcommand{\eq}{Eq.}

\newcommand{\fig}{Fig.}

\newcommand{\Refe}{Ref.}
\newcommand{\Refs}{Refs.}

\newcommand{\equ}[1]{\eq~(\ref{equ:#1})}
\newcommand{\figu}[1]{\fig~\ref{fig:#1}}

\newcommand{\bi}{\begin{itemize}}
\newcommand{\ei}{\end{itemize}}

\newcolumntype{T}{>{\centering\arraybackslash}m{1.5cm}}
\newcolumntype{L}{>{\centering\arraybackslash}m{2cm}}

\begin{document}
 
\title{Using High-Energy Neutrinos As Cosmic Magnetometers}

\author{Mauricio Bustamante}
\email{mbustamante@nbi.ku.dk}
\thanks{ORCID: \href{http://orcid.org/0000-0001-6923-0865}{0000-0001-6923-0865}}
\affiliation{Niels Bohr International Academy \& DARK, Niels Bohr Institute,\\University of Copenhagen, 2100 Copenhagen, Denmark}

\author{Irene Tamborra}
\email{tamborra@nbi.ku.dk}
\thanks{ORCID: \href{http://orcid.org/0000-0001-7449-104X}{0000-0001-7449-104X}}
\affiliation{Niels Bohr International Academy \& DARK, Niels Bohr Institute,\\University of Copenhagen, 2100 Copenhagen, Denmark}

\date{September 2, 2020}

\begin{abstract}
Magnetic fields are crucial in shaping the non-thermal emission of the TeV--PeV neutrinos of astrophysical origin seen by the IceCube neutrino telescope.
The sources of these neutrinos are unknown, but if they harbor a strong magnetic field, then the synchrotron energy losses of the neutrino parent particles---protons, pions, and muons---leave characteristic imprints on the  neutrino energy distribution and its flavor composition. 
We use high-energy neutrinos as ``cosmic magnetometers'' to constrain the identity of their sources by placing limits on the strength of the magnetic field in them.
We look for evidence of synchrotron losses in public IceCube data: 6 years of High Energy Starting Events (HESE) and 2 years of Medium Energy Starting Events (MESE).  
In the absence of evidence, we place an upper limit of $10$~kG--$10$~MG (95\%~C.L.) on the average magnetic field strength of the sources.
\end{abstract}

\maketitle


\section{Introduction}

Magnetic fields are pivotal to the dynamics of high-energy astrophysical sources.  They help to launch and collimate outflows in relativistic jets, affect matter accretion processes, and aid angular momentum transport.  
Magnetic fields also play a crucial role in the emission of high-energy astrophysical neutrinos, gamma rays, and cosmic rays.
Although the sources of these particles are largely unknown, a fundamental requirement is that they must harbor a magnetic field capable of accelerating protons and charged nuclei to PeV energies or more\ \cite{Anchordoqui:2013dnh, Aloisio:2017qoo, Anchordoqui:2018qom, AlvesBatista:2019tlv}.
Some high-energy protons and nuclei escape as cosmic rays; others interact with surrounding matter and radiation to produce neutrinos and gamma rays.

The TeV--PeV neutrinos detected by the IceCube neutrino telescope\ \cite{Aartsen:2013bka, Aartsen:2013jdh, Aartsen:2014gkd, Aartsen:2015rwa, Aartsen:2016xlq} are especially powerful source tracers, due to their low chance of being stopped or deflected en route to Earth.  Yet, direct\ \cite{Bartos:2016wud, Aartsen:2016lir, Aartsen:2017wea, Mertsch:2016hcd,Aartsen:2018ywr} and indirect\ \cite{Murase:2013rfa, Murase:2015xka, Ando:2015bva, Silvestri:2009xb, Ahlers:2014ioa, Murase:2016gly, Dekker:2018cqu,Feyereisen:2016fzb}  searches have not provided conclusive evidence, save for two cases of probable identification\ \cite{IceCube:2018cha, Stein:2020xhk}.
Remarkably, the role of the source magnetic field provides us with a novel indirect search strategy.

\begin{figure}[t!]
 \centering
 \includegraphics[width=\columnwidth]{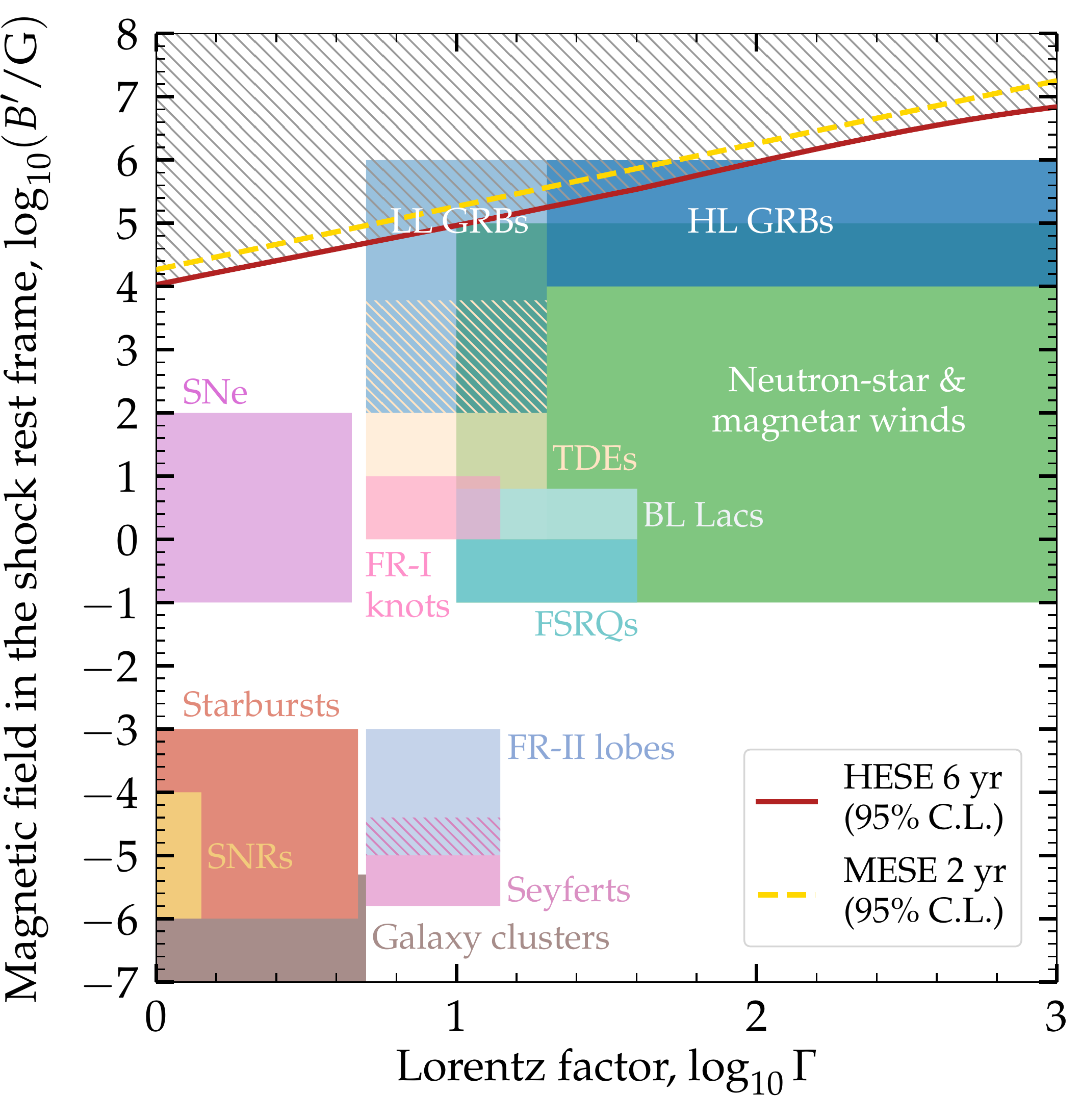}
 \caption{\label{fig:limits}Upper limit on the average magnetic field ($B^\prime$) as a function of the bulk Lorentz factor ($\Gamma$) of the astrophysical sources of TeV--PeV neutrinos, derived from 6 years of IceCube High Energy Starting Events\ \cite{Kopper:2015vzf, IC4yrHESEURL, Kopper:2017zzm} (HESE) and 2 years of Medium Energy Starting Events\ \cite{Aartsen:2014muf, IC2yrMESEURL} (MESE).  We include approximate ranges for candidate sources: neutron-star and magnetar winds\ \cite{Arons:2002yj, Bucciantini:2007hy, Mereghetti:2008je, Ptitsyna:2008zs}, low-\ \cite{Ghisellini:2006ng, Toma:2006iu} and high-luminosity\ \cite{Piran:2005qu, Ghirlanda:2017opl} gamma-ray bursts (LL GRBs, HL GRBs), blazars\ \cite{Ghisellini:2017ico} [flat spectrum radio quasars (FSRQs), BL Lacs], tidal disruption events\ \cite{Burrows:2011dn} (TDEs), starbursts\ \cite{Thompson:2006is, Ackermann:2012vca}, supernovae\ \cite{Murase:2018okz} (SNe), supernova remnants\ \cite{Wang:2007ya} (SNRs), galaxy clusters\ \cite{Brunetti:2014gsa, Voit:2004ah}, and active galactic nuclei (Seyfert\ \cite{Ptitsyna:2008zs, Paliya:2019oip}, FR-I\ \cite{Tjus:2014dna}, FR-II galaxies\ \cite{Tjus:2014dna}).  The average magnetic field strength is limited to be smaller than $10$~kG--$10$~MG.}
\end{figure}

We use TeV--PeV astrophysical neutrinos as ``cosmic magnetometers'' that constrain the average magnetic field of the neutrino sources.  Because candidate source classes span a wide range of magnetic field strengths, this constraint narrows down the identity of the sources.  

We look for imprints left by the magnetic field of the sources on the diffuse neutrino flux. 
On the one hand, the average magnetic field must be strong enough to accelerate protons up to PeV energies.  On the other hand, it cannot be too strong, or else proton energy losses via synchrotron radiation would lower the maximum proton energy and preclude the production of high-energy neutrinos.  Further, intermediate magnetic field strengths may induce synchrotron losses in secondary pions and muons that decay into neutrinos, affecting their energy spectrum and flavor composition.  We look for evidence of the interplay of these effects in public IceCube data.

Figure\ \ref{fig:limits} shows that our results limit the average magnetic field to be weaker than $10$~kG--$10$~MG.  This partially  disfavors low-luminosity GRBs as the main sites of TeV--PeV neutrino production.  Our work builds on and extends the  constraints on magnetic fields of a few MG found by \Refe\ \cite{Winter:2013cla} using the first 2 years of IceCube data, by using 4 more years of IceCube data, two different IceCube event samples, a significantly refined computation of event spectra, and a Bayesian statistical treatment.


\section{Energy losses via synchrotron radiation}

We consider a generic scenario that captures the key features common to high-energy neutrino source candidates.  Neutrinos are produced in an outflow of baryon-loaded material with a bulk Lorentz factor $\Gamma$ and magnetic field strength $B^\prime$.  To keep our scenario generic, we consider protons only; see, \eg, \Refs\ \cite{Boncioli:2016lkt, Biehl:2017zlw} for production scenarios including heavier nuclei and nuclear cascades.  Here and below, primed quantities are expressed in the rest frame of the neutrino production region, \ie, the shock rest frame; all other quantities are in the comoving frame of the production region or, after accounting for redshift effects, in the frame of the observer.

Sources accelerate protons via collisionless shocks in the magnetized outflow, up to a maximum energy $E_p^{\prime \max}$~\cite{Hillas:1985is}. To treat all candidate sources on an equal footing, we assume that $E_p^{\prime \max}$ is limited by proton synchrotron losses. However, alternative energy-loss mechanisms may be dominant in  some source classes\ \cite{Murase:2013ffa,Tamborra:2015qza,Fang:2014qva} or the flavor content may be affected in an energy-dependent fashion by oscillations in dense media\ \cite{Carpio:2020app,Razzaque:2004yv,Ando:2005xi}; we comment on source-specific features later.  Protons reach their maximum energy when two time scales become comparable: their acceleration time scale, $t_{\rm acc}^\prime = E_p^\prime/(\eta e B^\prime)$, where $e$ is the electron charge and $\eta$ is the acceleration efficiency, and their synchrotron energy-loss time scale, $t_{\rm sync}^\prime = 9 m_p^4/(4 e^4 B^{\prime 2} E_p^\prime)$, where $m_p$ is the proton mass.  In the comoving frame, this yields $E_p^{\max} \approx 2 \times 10^{11}~\Gamma ( \eta/B^\prime )^{1/2}$~GeV.  Since sources must be rather efficient accelerators, we fix $\eta = 1$ from here on, and comment later on how the neutrino emission changes with $\eta$.

The interactions of protons with matter and radiation produce secondary pions and muons that, upon decaying, generate neutrinos\ \cite{Stecker:1978ah}: $\pi^+ \to \mu^+ +  \nu_{\mu}$, followed by $\mu^+ \to e^+ + \nu_e + \bar{\nu}_\mu$, and the charge-conjugated processes~\cite{Kelner:2006tc,Kelner:2008ke}.  
On average, a pion receives 1/5 of the proton energy, each neutrino from muon decay receives 1/3 of the muon energy, and each final-state neutrino receives 1/4 of the pion energy. 
Following theory expectations, each source emits neutrinos distributed in energy as $E_\nu^{\prime 2}~dN_\nu / dE_\nu^\prime \propto E_\nu^ {\prime 2-\alpha_\nu} e^{-E_\nu^\prime/E_\nu^{\prime\max}}$, where $E_\nu^\prime$ is the neutrino energy.  
We assume that the maximum neutrino energy, $E_\nu^{\prime \max}$, and the spectral index, $\alpha_\nu$, are common to neutrinos and anti-neutrinos of all flavors.
Because each neutrino receives 1/20 of the parent proton energy, $E_\nu^{\max} = E_p^{\max}/20$ and is affected by synchrotron losses.  
Below, we show how synchrotron losses of the secondaries affect $\alpha_\nu$.

For the secondary pions, synchrotron losses are significant if they occur within a time scale shorter than the pion decay time, $t_{\rm dec}^\prime = \tau_\pi E_\pi^\prime / m_\pi$, where $E_\pi^\prime$, $\tau_\pi$, and $m_\pi$ are the pion energy, lifetime, and mass.  
In the comoving frame, this occurs at neutrino energies above $E_{\nu, \pi}^{\rm sync} \approx 3 \times 10^{10} (\Gamma/B^{\prime})$~GeV.
By analogous arguments, for the secondary muons,  synchrotron losses are significant at energies above $E_{\nu, \mu}^{\rm sync} \approx 2 \times 10^{9} (\Gamma/B^{\prime})$~GeV.

Below $E_{\nu, \pi}^{\rm sync}$, $\alpha_\nu$ is solely determined  by the parent proton and photon spectra.  In lieu of detailed modeling, we parametrize it as $\alpha_\nu(E_\nu < E_{\nu,\pi}^{\rm sync}) = \gamma$, where $\gamma \in [2,3]$ is a free parameter whose value we vary later.  
At $E_{\nu, \mu}^{\rm sync}$, the neutrino spectrum coming from the decay of muons steepens by $\sim$$E_\nu^{-2}$, so these neutrinos become sub-dominant and the flux is mainly from neutrinos produced in the direct decay of pions.  
As a result, the flavor composition of the emitted neutrinos, \ie, the fraction $f_{\alpha, {\rm S}}$ ($\alpha = e, \mu, \tau$) of neutrinos plus anti-neutrinos of each flavor, changes from that of the full pion decay chain, $\left( f_{e, {\rm S}} , f_{\mu, {\rm S}} , f_{\tau, {\rm S}} \right) = \left( 1/3, 2/3, 0 \right)$, at $E_\nu < E_{\nu, \mu}^{\rm sync}$, to that coming  from the direct pion decay only, $\left( 0, 1, 0 \right)$, at $E_\nu \geq E_{\nu, \mu}^{\rm sync}$.  At even higher energies $E_\nu \geq E_{\nu, \pi}^{\rm sync}$, the neutrino spectrum from pion decays itself steepens by $\sim$$E_\nu^{-2}$, so $\alpha_\nu(E_\nu \geq E_{\nu, \pi}^{\rm sync}) = \gamma+2$.


\section{Diffuse flux of high-energy neutrinos}

The luminosity of $\nu_\alpha + \bar{\nu}_\alpha$ that reaches Earth from a single source located at redshift $z$, in the frame of the observer, is
\begin{eqnarray}
 J_{\nu_\alpha}&&(E_\nu, z, \gamma, \Gamma, B^\prime)
 \propto f_{\alpha, \oplus}(E_\nu (1+z), \Gamma, B^\prime) \\ \nonumber
 && \times~ [E_\nu (1+z)]^{2-\alpha_\nu(E_\nu, \gamma, \Gamma, B^\prime)} 
 ~ e^{ - E_\nu(1+z) / E_{\nu}^{\max}(\Gamma, B^\prime) } \; .
\end{eqnarray}
Because flavors mix, neutrino flavor conversions en route to Earth change the flavor composition into $f_{\alpha, \oplus} = \sum_{\beta=e,\mu,\tau} P_{\beta\alpha} f_{\beta, {\rm S}}$, where $P_{\beta\alpha}$ is the average $\nu_\beta \to \nu_\alpha$ conversion probability\ \cite{Pakvasa:2008nx}.  To compute it, we fix the mixing parameters to their best-fit values from the recent {\sc NuFit} 4.1 global fit to neutrino oscillation data\ \cite{Esteban:2018azc, NuFit_4.1}, assuming normal neutrino mass ordering.  
The flavor composition changes from $(f_{e, \oplus}, f_{\mu, \oplus}, f_{\tau, \oplus}) \approx (1/3, 1/3, 1/3)$ at $E_\nu < E_{\nu, \mu}^{\rm sync}$ to roughly $(1/5, 2/5, 2/5)$ at $E_\nu \geq E_{\nu, \mu}^{\rm sync}$. 

To compute the diffuse flux of $\nu_\alpha + \bar{\nu}_\alpha$, $\Phi_{\nu_\alpha}$, we integrate the contribution from all sources up to redshift $z_{\max} = 4$; sources at higher redshifts contribute negligibly.  To describe a variety of candidate source classes, we adopt the following parametrization for the source density: $\rho \propto (1+z)^m$ up to $z_c \equiv 1.5$, and $\rho  \propto (1+z_c)^m$ at $z > z_c$.  Later, we let the value of $m$ float.  Thus, the diffuse energy flux is 
\begin{eqnarray}
 && E_\nu^2 \Phi_{\nu_\alpha} ( E_\nu, \gamma, m, \Gamma, B^\prime) \\ \nonumber
 && \qquad \propto
 \int_0^{z_{\max}} dz
 \frac{\rho(z, m)}{h(z)(1+z)^2}
 J_{\nu_\alpha}(E_\nu, z, \gamma, \Gamma, B^\prime) \;,
\end{eqnarray}
where $h(z) \equiv [\Omega_\Lambda + (1+z)^3 \Omega_m]^{1/2}$ is the adimensional Hubble parameter, and $\Omega_\Lambda = 0.685$ and $\Omega_m = 0.315$ are the energy densities of vacuum and matter\ \cite{Aghanim:2018eyx, Tanabashi:2018oca}.   

We assume that all of the contributing sources have the same values of $\gamma$, $\Gamma$, and $B^\prime$.  In reality, these parameters likely follow distributions that are presently unknown. By assuming values that are common to all sources, we aim to constrain their population-averaged values.

In summary, if the average $B^\prime$ is large, the synchrotron losses of protons and secondaries may visibly affect the spectral index $\alpha_\nu$, flavor composition $f_{\alpha, \oplus}$, and maximum energy $E_\nu^{\max}$ of the diffuse neutrino flux\ \cite{Kashti:2005qa, Kachelriess:2007tr, Lipari:2007su, Baerwald:2010fk, Winter:2011jr}.  These features are softened and spread out in energy by the redshift distribution of the sources.

\begin{figure}[t!]
 \centering
 \includegraphics[width=\columnwidth, height=\columnwidth]{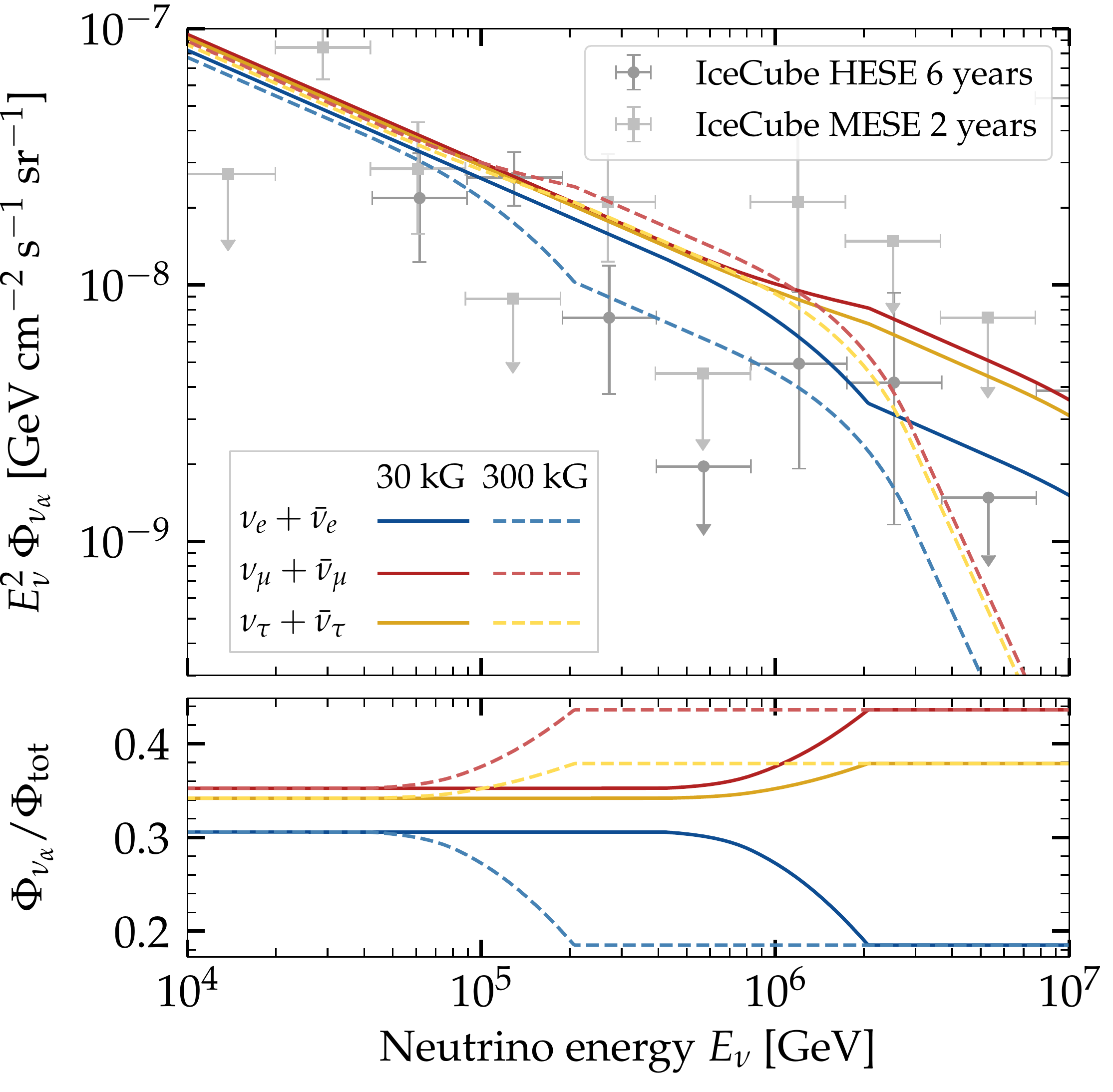}
 \caption{\label{fig:fluxes}{{\it Top:} Sample diffuse astrophysical neutrino fluxes at Earth, as a function of the neutrino energy, for two illustrative values of the magnetic field strength of the sources, $B^\prime$.  In this plot only, the flux parameters are fixed to illustrative values: $\gamma = 2.50$, $\Gamma \approx 32$, $m = 1.5$, and the normalization of each flux, at $E_\nu = 100$~TeV, to $3 \times 10^{-8}$~GeV~cm$^{-1}$~s$^{-1}$~sr$^{-1}$.  Data points show the spectra of the IceCube HESE\ \cite{Kopper:2015vzf, IC4yrHESEURL, Kopper:2017zzm} and MESE\ \cite{Aartsen:2014muf, IC2yrMESEURL} event samples.  {\it Bottom:} Flavor composition $\Phi_{\nu_\alpha}/\Phi_{\rm tot}$ ($\alpha = e, \mu, \tau$) of the diffuse flux, where $\Phi_{\rm tot} \equiv \sum_\alpha \Phi_{\nu_\alpha}$.  The spectral breaks and changes to the flavor composition are due to the synchrotron cooling of muons and pions.}}
 \label{fig:diffuse}
\end{figure}

Figure~\ref{fig:diffuse} shows sample fluxes for two choices of $B^\prime$.  The change to the flavor composition due to muon synchrotron cooling is prominent in both fluxes, at $E_\nu \approx 2$~PeV and 200~TeV for $B^\prime = 30$~kG and 300~kG, respectively.  For the flux with $B^\prime = 300$~kG, the spectral softening due to pion synchrotron cooling is also visible around 3~PeV. 
In \figu{diffuse}, the flux dampening due to proton synchrotron cooling occurs at energies higher than shown.
This is true for viable neutrino source candidates, which must be efficient accelerators: as long as $\eta \gtrsim 0.01$, proton cooling becomes important only after muon cooling does, provided $B^\prime$ is at least a few G. 

\subsection{Neutrino propagation through the Earth}

Upon reaching Earth, high-energy neutrinos propagate from its surface, through its interior, and up to the South Pole, where IceCube is located.  Neutrino-nucleon 
interactions along the way modify the neutrino flux.  At these energies, neutrinos deep-inelastic scatter off of nucleons.  Charged-current (CC) interactions ($\nu_\alpha + N \to \alpha + X$, where $X$ are final-state hadrons) remove neutrinos from the flux.   Neutral-current (NC) interactions ($\nu_\alpha + N \to \nu_\alpha + X$) redistribute neutrinos from high to low energies. 

To compute the neutrino flux that reaches IceCube, we adopt the Preliminary Reference Earth Model\ \cite{Dziewonski:1981xy} for the matter density inside the Earth and assume that matter is isoscalar, \ie, made up of equal numbers of protons and neutrons.  At these energies, there are no matter-driven flavor transitions\ \cite{Wolfenstein:1977ue, Mikheev:1986gs}.
The flux at IceCube is different for different propagation directions, flavors, and for neutrinos and anti-neutrinos.
We use {\sc nuSQuIDS}\ \cite{Delgado:2014kpa, SQuIDS, NuSQuIDS} to compute the fluxes of $\nu_\alpha$ and $\bar{\nu}_\alpha$, astrophysical and atmospheric (see below), that reach IceCube.


\subsection{Atmospheric backgrounds}

The main backgrounds to our analysis are high-energy atmospheric neutrinos and muons produced in the interaction of cosmic rays in the atmosphere of the Earth.  They are especially important below 100~TeV\ \cite{Beacom:2004jb}.  In IceCube, the contamination of atmospheric neutrinos is mitigated by using part of the detector as a self-veto\ \cite{Schonert:2008is, Gaisser:2014bja, Aartsen:2014muf, Arguelles:2018awr}, and the contamination of atmospheric muons, by using a surface array of water-Cherenkov tanks as veto\ \cite{Aartsen:2013jdh, Tosi:2019nau}.  In our analysis, we account for these backgrounds and vetoes.

For atmospheric neutrinos, we use the same state-of-the-art tools used by the IceCube Collaboration: {\sc MCEq}\ \cite{Fedynitch:2015zma, MCEq} to generate neutrino fluxes produced in the decay of pions and kaons, and {\sc nuVeto}\ \cite{Arguelles:2018awr, nuVeto} to compute the flux reduction due to the self-veto for HESE events.  For MESE events, the self-veto is already included in the effective detector area; see Appendix\ \ref{app:mese_events}.  We do not consider prompt atmospheric neutrinos produced in the decay of charmed mesons, since they have not been observed and are subject to stringent upper limits\ \cite{Aartsen:2015knd}.  

For atmospheric muons, we approximate the flux that reaches IceCube, after the surface veto, following the procedure in \Refe\ \cite{Palomares-Ruiz:2015mka} for HESE events, and extending it for MESE events, as detailed in Appendix\ \ref{app:mese_muon}.


\subsection{High-energy neutrino detection}

In IceCube, neutrinos are detected when they scatter off of nucleons in the Antarctic ice and trigger particle showers that emit Cherenkov light that is collected by photomultipliers buried 1.5--2.5~km underground.  From the amount of light collected and from its spatial and temporal profiles, IceCube infers the energy deposited, $E_{\rm dep}$, and the neutrino arrival direction, $\cos \theta_z$, where $\theta_z$ is the zenith angle measured from the South Pole.  

We focus on ``starting'' events, where the neutrino interaction occurs inside the instrumented volume, so that $E_{\rm dep}$ is close to the energy of the interacting neutrino.  In the TeV--PeV range, events are predominantly ``showers,'' roughly spherical light profiles triggered mainly by $\nu_e$ and $\nu_\tau$, and ``tracks,'' elongated light profiles made by final-state muons, triggered mainly by $\nu_\mu$.  

For an incoming flux of astrophysical or atmospheric neutrinos along a given direction,
we forecast the spectra $dN/dE_{\rm dep}$ of showers and tracks following \Refe\ \cite{Palomares-Ruiz:2015mka}; see Appendix\ \ref{app:hese_events}.  We account for differences in deposited energy for different flavors, CC vs.~NC, and decay channels of final-state particles, and for the $\sim$$13\%$ detector energy resolution.


\section{Statistical analysis}

We compare forecasted event spectra computed with different values of the flux parameters against two public IceCube event samples: 6 years of High Energy Starting Events (HESE)\ 
\cite{Kopper:2015vzf, IC4yrHESEURL, Kopper:2017zzm} and 2 years of Medium Energy Starting Events (MESE)\ \cite{Aartsen:2014muf, IC2yrMESEURL}.
The HESE sample has 58 showers and 22 tracks with $E_{\rm dep} =$~18~TeV--2~PeV.  The MESE sample has 278 showers and 105 tracks with $E_{\rm dep} =$~330~GeV--1.3~PeV.  The samples provide $E_{\rm dep}$ and $\cos\theta_z$ for each event.  A few events are common to both samples, but since they cannot be singled out, we treat the samples separately to avoid double-counting.

In the HESE sample, below a few tens of TeV, roughly half of the events are of atmospheric origin and half of astrophysical origin; above, they are mostly of astrophysical origin\ \cite{Kopper:2017zzm}.  The MESE sample extends to lower energies, where the atmospheric contamination is higher. 
We restrict MESE events to $E_{\rm dep} \geq 20$~TeV in order to avoid a dominant atmospheric contamination that, according to our tests, would otherwise skew our results.  This is the same energy region of validity of the astrophysical neutrino component in the IceCube MESE analysis\ \cite{Aartsen:2014muf}.

We adopt a Bayesian approach to search for evidence of synchrotron cooling and constrain $B^\prime$.  For a sample of $N_{\rm obs}$ IceCube events, our likelihood function is
\begin{eqnarray}
 \label{equ:likelihood}
 && \mathcal{L}(\gamma, m, \Gamma, B^\prime, N_{\rm ast}, N_{\rm atm}, N_\mu) 
 \\  \nonumber
 && 
 =
 e^{-N_{\rm ast}-N_{\rm atm}-N_\mu}
 \prod_{i=1}^{N_{\rm obs}} \mathcal{L}_i(\gamma, m, \Gamma, B^\prime, N_{\rm ast}, N_{\rm atm}, N_\mu) \;,
\end{eqnarray}
where $N_{\rm ast}$, $N_{\rm atm}$, and $N_\mu$ are, respectively, the number of events due to astrophysical neutrinos, atmospheric neutrinos, and atmospheric muons.  The partial likelihood for the $i$-th event compares the odds of it being due to the different fluxes:
$\mathcal{L}_i = N_{\rm ast} \mathcal{P}_{i, {\rm ast}}(\gamma, m, \Gamma, B^\prime) + N_{\rm atm} \mathcal{P}_{i, {\rm atm}} + N_{\mu} \mathcal{P}_{i, \mu}$, where $\mathcal{P}_{i, {\rm ast}}$, $\mathcal{P}_{i, {\rm atm}}$, and $\mathcal{P}_{i, \mu}$ are the probability distribution functions for this event to have been generated by astrophysical neutrinos, atmospheric neutrinos, and atmospheric muons.  For astrophysical neutrinos, $\mathcal{P}_{i, {\rm ast}} = (dN_i/dE_{\rm dep}) / \int dE_{\rm dep} (dN_i/dE_{\rm dep})$, where the event spectrum is computed using the flux along the direction of the $i$-th event, and similarly for atmospheric neutrinos.  The procedure differs slightly for HESE and MESE, and for muons; see Appendix\ \ref{app:event_spectra}.

We maximize the likelihood separately for the HESE and MESE samples.  For $N_{\rm ast}$, $N_{\rm atm}$, $N_\mu$, and $\gamma$ (only for HESE), we adopt informed priors based on the HESE\ \cite{Kopper:2017zzm} and MESE\ \cite{Aartsen:2014muf} analyses by IceCube.  For $m$, $\log_{10} \Gamma$, and $\log_{10}(B^\prime/{\rm G})$, we adopt wide uniform priors to avoid introducing bias.  See Appendix\ \ref{app:likelihood} for details about the priors.  To maximize the likelihood, we use the efficient Bayesian sampler {\sc MultiNest}\ \cite{Feroz:2007kg, Feroz:2008xx, Feroz:2013hea, Buchner:2014nha}.  We quantify the preference for synchrotron-loss features via the Bayes factor, $K \equiv Z_{\rm signal}/Z_{\rm null}$, that compares the evidence, or marginalized likelihood, in favor of their presence, $Z_{\rm signal}$, over their absence, $Z_{\rm null}$.  Appendix\ \ref{app:stat_analysis} contains details of the statistical analysis.

\begin{table}
 \begin{ruledtabular}
  \caption{\label{tab:fit_results}
  Allowed marginalized ranges of the average Lorentz factor, $\Gamma$, and magnetic field strength, $B^\prime$, of the sources of high-energy astrophysical neutrinos, obtained using the public IceCube 6-year HESE\ \cite{Kopper:2015vzf, IC4yrHESEURL, Kopper:2017zzm} and 2-year MESE\ \cite{Aartsen:2014muf, IC2yrMESEURL} event samples.  The preference for synchrotron-loss features in the data is insignificant; see the main text.}
  \centering
  \renewcommand{\arraystretch}{1.1}
  \begin{tabular}{ccccc}
   \multirow{2}{*}{Parameter}     & \multicolumn{2}{c}{HESE}               & \multicolumn{2}{c}{MESE ($E_{\rm dep} > 20$~TeV)} \\
                                  & Mean $\pm 1\sigma$ & 95\%~C.L.         & Mean $\pm 1\sigma$ & 95\%~C.L. \\
   \hline
   $\log_{10} \Gamma$             & $1.78 \pm 0.86$       & $[0.22, 2.86]$ & $1.16 \pm 0.90$       & $[0.85, 6.75]$ \\
   $\log_{10}(B^\prime/{\rm G})$  & $2.96 \pm 1.78$       & $[0.23, 5.97]$ & $4.82 \pm 1.20$       & $[0.10, 2.82]$ 
  \end{tabular}
 \end{ruledtabular}
\end{table}


\section{Results}

Table\ \ref{tab:fit_results} shows the allowed ranges of $\Gamma$ and $B^\prime$ that result from our analysis.  Both event samples prefer values of $B^\prime$ that push the synchrotron-loss features to energies 
beyond a few PeV, past the energies covered by the samples.  
For both samples, $\log_{10} K \approx 0.3$, \ie, the evidence for synchrotron-loss features in the data is insignificant\ \cite{Jeffreys:1939xee}.
Hence, we place upper limits on $B^\prime$.

Figure~\ref{fig:limits} shows our limits on $B^\prime$ as a function of $\Gamma$, after marginalizing over all other likelihood parameters.  The limits are isocontours of $E_{\nu, \mu}^{\rm synch}$ and $E_{\nu, \pi}^{\rm synch}$ and, in agreement with \Refe\ \cite{Winter:2013cla}, we find that they are predominantly driven  by the absence of a spectral break, rather than by the absence of a change in flavor composition, to which there is currently little sensitivity.  The limits are similar for both samples because, after selecting MESE events with $E_{\rm dep} \geq 20$~TeV, both samples cover roughly the same $E_{\rm dep}$ range.  The MESE limit is worse because there are 26 fewer MESE than HESE events. 

Our results disfavor a predominant origin of the TeV--PeV neutrinos in astrophysical sources with average magnetic field stronger than $\sim$$10(1+\Gamma)$~kG, \ie, $10$~kG--$10$~MG approximately.  This partially includes low-luminosity GRBs\ \cite{Gupta:2006jm, Murase:2006mm, Kashiyama:2012zn, Murase:2013ffa, Tamborra:2015qza, Senno:2015tsn, Zhang:2017moz}.  Fast radio bursts (FRBs) have been considered as potential neutrino sources, but their feasibility as such depends on their origin, which is currently subject of intense investigation.  So far, direct searches have found no evidence for FRBs as high-energy neutrino sources\ \cite{Fahey:2016czk, Aartsen:2017zvw, Albert:2018euo, Aartsen:2019wbt, Metzger:2020byf}.  If FRBs are connected to magnetars\ \cite{Li:2013kwa, Dey:2016psn, Andersen:2020hvz, Ridnaia:2020gcv, Bochenek:2020zxn, Tavani:2020adq}, then our limits would disfavor relativistic outflows in FRBs as regions of copious production of high-energy neutrinos.  Our results also disfavor the non-thermal emission of high-energy neutrinos from the crusts of magnetars and neutron stars\ \cite {Zhang:2002xv}, with $B^\prime \approx 10^{14}$--$10^{15}$~G, but these sites are not expected to be efficient hadronic accelerators.

Sources with an intermediate field of $10$~kG--$1$~MG---high-luminosity GRBs\ \cite{Waxman:1997ti, Murase:2005hy, Hummer:2011ms, Zhang:2012qy, Liu:2012pf, Baerwald:2014zga, Tamborra:2015qza, Tamborra:2015fzv, Biehl:2017zlw},
blazars\ \cite{Stecker:1991vm, Atoyan:2001ey, Murase:2011cy, Padovani:2014bha, Kimura:2014jba, Padovani:2015mba, Cerruti:2018tmc, Liu:2018utd, Palladino:2018lov,  Keivani:2018rnh, Murase:2018iyl, Gao:2018mnu, Rodrigues:2018tku, Petropoulou:2019zqp}, pulsar and magnetar winds~\cite{Arons:2002yj, Ptitsyna:2008zs, Fang:2014qva, Fang:2017tla, Fang:2015xhg}---remain viable candidates according to our analysis. 
These are electromagnetically luminous and relatively abundant sources; at face value, they are ideal neutrino emitters.
However, they are strongly constrained by dedicated searches: the contribution of high-luminosity GRBs and blazars is restricted to be less than 2\%\ \cite{Aartsen:2017wea} and 15\%\ \cite{Smith:2020oac, Aartsen:2016lir} of the diffuse flux, respectively.  

Sources with a weak field---non-blazar AGN\ \cite{Anchordoqui:2004eu, Stecker:2005hn, AlvarezMuniz:2004uz, Stecker:2013fxa, Tjus:2014dna, Hooper:2018wyk, Smith:2020oac}, TDEs\ \cite{Dai:2016gtz, Lunardini:2016xwi, Biehl:2017hnb, Winter:2020ptf, Murase:2020lnu}, starburst galaxies\ \cite{Loeb:2006tw, Stecker:2006vz, Thompson:2006np, He:2013cqa, Tamborra:2014xia, Chang:2014hua, Hooper:2018wyk, Peretti:2019vsj}, supernovae\ \cite{Horiuchi:2007xi, Murase:2017pfe, Petropoulou:2017ymv, Senno:2017vtd, Murase:2018okz, Esmaili:2018wnv, Fang:2020bkm}, supernova remnants\ \cite{Wang:2007ya, Villante:2008qg, Liu:2013wia, Chakraborty:2015sta}, and galaxy clusters\ \cite{Dar:1995gk, Berezinsky:1996wx, DeMarco:2005va, Kotera:2009ms, Zandanel:2014pva, Senno:2015tra}---remain viable candidates in a broader sense and may account for a large fraction of the diffuse neutrino flux.  The neutrino emission from these sources is largely unconstrained by direct searches due to their high abundance and low luminosity per source\ \cite{Silvestri:2009xb, Murase:2016gly,Mertsch:2016hcd, Capel:2020txc,Fang:2020rvq}.  

While we have derived our results assuming that the sources emit neutrinos with a power-law energy distribution, they are valid within a more general framework.  We have repeated our analysis using a broken power-law energy distribution where the spectral index $\alpha_\nu$ steepens by $\Delta \alpha_\nu$ after a break energy $E_{\nu, {\rm br}}^\prime$.  This steepening mimics a production scenario where neutrinos inherit the peaked shape of a parent photon spectrum or the pile-up of low-energy neutrinos produced by the decay of synchrotron-cooled muons; see, \eg, \Refs\ \cite{Waxman:1998yy, Mannheim:1998wp, Mucke:1999yb, Kelner:2008ke, Hummer:2010vx}.  Also in this case, we find no evidence of synchrotron-loss features in the HESE and MESE samples.  The resulting upper limits on $B^\prime$ are very similar to those shown in \figu{limits}.


\section{Summary and outlook}

The magnetic field of the sources that populate the high-energy Universe remain enigmatic.  At the same time, one of the main goals of neutrino astronomy is to identify the sources of the TeV--PeV neutrinos seen by IceCube.   We have introduced a new way to address these two long-standing questions, by using high-energy neutrinos as cosmic magnetometers.  

We look in the diffuse flux of TeV--PeV astrophysical neutrinos for imprints left by the magnetic field of their sources on the energy spectrum and flavor composition.  These imprints originate in the energy losses via synchrotron radiation of the protons, pions, and muons that produce the neutrinos.  We find no evidence in 6 years of IceCube high-energy events (HESE) and 2 years of medium-energy events (MESE).  Thus, we constrain the average magnetic field strength of the neutrino sources to be smaller than $10$~kG--$10$~MG.  Consequently, we partially disfavor low-luminosity GRBs as the predominant sources of TeV--PeV neutrinos, but sources with a weak magnetic field---AGN, TDEs, SNe, SNRs, starburst galaxies, magnetar and neutron star winds, and galaxy clusters---remain viable candidates.

Because we use a generic model of neutrino production, our results apply to a wide range of candidate source classes.  Future work may explore source- and population-dependent modeling in order to boost the sensitivity to specific source classes.  Further refinements include a detailed treatment of the source physics and cosmic-ray acceleration, the interaction not only of protons, but also of nuclei, additional neutrino-production channels, and non-synchrotron losses; see, \eg, \Refs\ \cite{Anchordoqui:2007tn, Hummer:2010vx, Petropoulou:2014awa, Boncioli:2016lkt, Morejon:2019pfu,Fang:2014qva}.  Further, the acceleration of secondary muons, pions, and kaons, which we have neglected, could play an important role in mitigating synchrotron cooling in astrophysical environments with efficient particle diffusion\ \cite{Murase:2011cx, Winter:2014tta}.

In the future, larger detectors, like the planned IceCube-Gen2~\cite{Aartsen:2020fgd}, will be able to detect neutrinos at a higher rate, at energies beyond the PeV scale, and will have improved sensitivity to changes in flavor composition with energy\ \cite{Bustamante:2015waa}.  This will allow to place tighter bounds on the source magnetic field and probe the existence of synchrotron-loss features at higher energies, \ie, of magnetic fields weaker than the ones that we are currently sensitive to.


\acknowledgments

We thank Ke Fang, Kohta Murase, Foteini Oikonomou, and Walter Winter for helpful discussions and feedback on the manuscript.  This project was supported by the Villum Foundation (Project No.~13164), the Carlsberg Foundation (CF18-0183), the Knud H{\o}jgaard Foundation, and the Deutsche Forschungsgemeinschaft through Sonderforschungbereich SFB~1258 ``Neutrinos and Dark Matter in Astro- and Particle Physics'' (NDM).


\appendix

\section{Overview of the calculation of event spectra at IceCube}
\label{app:event_spectra}

\subsection{HESE events}
\label{app:hese_events}

To compute the differential spectrum $dN/dE_{\rm dep}$ of HESE events at IceCube, we follow the detailed procedure from \Refe\ \cite{Palomares-Ruiz:2015mka}.  The spectra of showers and tracks are computed separately, accounting in each case for the contributing interactions of all flavors of neutrinos and anti-neutrinos.  The event rates are computed from first principles, \ie, from the effective IceCube mass and the deep-inelastic-scattering neutrino-nucleon differential cross section for $\nu_\alpha$ and $\bar{\nu}_\alpha$, based on the CTEQ14 parton distribution functions\ \cite{Dulat:2015mca}.  The procedure accounts for differences in deposited energy for different flavors, CC vs.~NC, and decay channels of final-state particles, and for the $\sim$$13\%$ detector energy resolution.  We defer to \Refe\ \cite{Palomares-Ruiz:2015mka} for the explanation of the full procedure and to Appendix A in \Refe\ \cite{Bustamante:2020mep} and Appendix C in \Refe\ \cite{Bustamante:2020niz} for an overview.


\subsection{MESE events}
\label{app:mese_events}

To compute the spectrum of MESE events, we use the IceCube effective area, $A_{{\rm eff}}^{s,t}(E_\nu, E_{\rm rec}, \cos \theta_z, \cos \theta_{z, {\rm rec}})$, provided by the IceCube Collaboration for its 2-year MESE analysis\ \cite{Aartsen:2014muf, IC2yrMESEURL}, as a function of true neutrino energy $E_\nu$, reconstructed (\ie, deposited) energy $E_{\rm rec}$, true neutrino direction $\cos \theta_z$, and reconstructed direction $\cos \theta_{z, {\rm rec}}$.  The effective area is provided separately for each neutrino species $s = \nu_e, \bar{\nu}_e, \nu_\mu, \bar{\nu}_\mu, \nu_\tau, \bar{\nu}_\tau$, and topology $t = $~sh (shower), tr (track).  It includes the effects of the MESE self-veto and of the mapping between true and reconstructed quantities.  

In general, given diffuse neutrino fluxes $\Phi_s$, the number of detected events at IceCube after a time $T$, at a given reconstructed energy and direction is
\begin{widetext}
\begin{equation}
 \label{equ:mese_events_general}
 N_t(E_{\rm rec}, \cos \theta_{z,{\rm rec}})
 =
 2 \pi T 
 \sum_s
 \int d E_\nu
 \int_{-1}^{+1} d \cos \theta_z
 A_{{\rm eff}}^{s,t}(E_\nu, E_{\rm rec}, \cos \theta_z, \cos \theta_{z, {\rm rec}})
 \Phi_s(E_\nu, \cos \theta_z) \;.
\end{equation}
\end{widetext}
The sum over $s$ adds the contribution of all flavors of neutrinos and anti-neutrinos that make showers or tracks.

For our analysis, we make two modifications to this expression.  First, because the effective area is binned in all of its input parameters, we change the integrals in \equ{mese_events_general} for sums over bins.  To do this, we write the effective area in the $j$-th bin of $E_{\rm rec}$, the $k$-th bin of $\cos \theta_{z,{\rm rec}}$, the $l$-th bin of $E_\nu$, and the $m$-th bin of $\cos \theta_z$ as $A_{{\rm eff},jklm}^{s,t}$.  Second, because there are only two bins of $\cos \theta_{z, {\rm rec}}$ provided ([0.2,1.0] for downgoing events and [-1,0.2] for upgoing events) we assume that $\cos \theta_{z, {\rm rec}} \approx \cos \theta_z$, \ie, that the reconstructed direction closely follows the neutrino direction.  This means that, in our computation, $m = k$ always. Thus, if the requested values of $E_{\rm rec}$ and $\cos \theta_{z,{\rm rec}}$ on the left-hand side of \equ{mese_events_general} are contained, respectively, inside their $j^\ast$-th and $k^\ast$-th bins, then \equ{mese_events_general} simplifies to 
\begin{widetext}
\begin{equation}
 \label{equ:mese_events_simplified}
 N_t(E_{\rm rec}, \cos \theta_{z,{\rm rec}})
 =
 2 \pi T 
 \sum_s
 \sum_{l=1}^{N_{E_\nu}}
 A_{{\rm eff},j^\ast k^\ast l k^\ast}^{s,t}
 \int_{E_{\nu, l}^{\min}}^{E_{\nu, l}^{\max}} d E_\nu
 \Phi_s(E_\nu, \cos \theta_{z,{\rm rec}}) \;,
\end{equation}
\end{widetext}
where $N_{E_\nu}$ is the number of bins of $E_\nu$, and $E_{\nu, l}^{\min}$ and $E_{\nu, l}^{\max}$ are the minimum and maximum energies in the $l$-th bin of $E_\nu$.

After this, the statistical procedure described in the main text proceeds similarly for MESE and HESE events.  There are only two differences.  The first one is in the computation of the spectra of atmospheric muons, which we describe in Appendix\ \ref{app:mese_muon}.  The second one is in the computation of the astrophysical and atmospheric probability distribution functions, $\mathcal{P}_{i,{\rm ast}}$ and $\mathcal{P}_{i,{\rm atm}}$, for the $i$-th MESE event.  The probability distribution functions are computed similarly as for HESE events, but using the total event rate, \equ{mese_events_simplified}, instead of the differential event rate (see the main text).  For astrophysical neutrinos, this is 
\begin{equation}
 \mathcal{P}_{i,{\rm ast}} 
 = 
 \frac { N_t(E_{{\rm rec},i}, \cos \theta_{z,{\rm rec},i}) } 
       { \sum_{j=1}^{N_{E_{\rm rec}}} N_t(E_{{\rm rec},j}, \cos \theta_{z,{\rm rec},i}) } \;,
\end{equation}
where $t$ depends on whether this event is a shower or track, $N_t$ is computed using the astrophysical neutrino fluxes, $E_{{\rm rec},i}$ and $\cos \theta_{z,{\rm rec},i}$ are the deposited energy and direction of the $i$-th event, and $N_{E_{\rm rec}}$ is the number of bins of $E_{\rm rec}$.  The calculation is analogous for $\mathcal{P}_{i,{\rm atm}}$, changing only the astrophysical fluxes for the atmospheric fluxes in \equ{mese_events_simplified}.


\subsection{Atmospheric muon background in the HESE and MESE samples}
\label{app:mese_muon}

To compute the spectra of atmospheric muons that reaches IceCube, we follow the procedure from \Refe\ \cite{Palomares-Ruiz:2015mka}, which was designed for HESE events.  We extend its application also to MESE events.  

Instead of simulating the propagation of muons through the Earth and applying a surface veto ourselves to mitigate their contamination, the procedure in \Refe\ \cite{Palomares-Ruiz:2015mka} directly computes the distribution $dN_\mu/dE_{\rm dep} \propto E_{\rm dep}^{-\gamma_\mu}$ of muons that pass the veto.  The value of the spectral index $\gamma_\mu$ varies for HESE versus MESE events.  To compute it, we compare the number of passing muons below and above a certain deposited energy $E_{\rm dep}^\ast$, respectively,
\begin{eqnarray}
 N_\mu(< E_{\rm dep}^\ast) 
 &=& \frac{1}{1-\gamma_\mu} \left[ (E_{\rm dep}^\ast)^{1-\gamma_\mu} - (E_{\rm dep}^{\rm min})^{1-\gamma_\mu} \right] \;, \\
 N_\mu(\geq E_{\rm dep}^\ast) 
 &=& \frac{-1}{1-\gamma_\mu} (E_{\rm dep}^\ast)^{1-\gamma_\mu} \;,
\end{eqnarray}
where $E_{\rm dep}^{\rm min}$ is the minimum deposited energy in the event sample.  With this, the spectral index is
\begin{equation}
 \gamma_\mu 
 =
 1 - \ln\left( 1 + \frac{N_\mu(< E_{\rm dep}^\ast)}{N_\mu(\geq E_{\rm dep}^\ast)} - \frac{E_{\rm dep}^{\rm min}}{E_{\rm dep}^\ast} \right) \;.
\end{equation}

For HESE events, we compute $\gamma_\mu$ using the muon rates reported in the 3-year IceCube HESE analysis\ \cite{Aartsen:2014gkd}.  Table IV in \Refe\ \cite{Aartsen:2014gkd} reports 8 passing muons below $E_{\rm dep}^\ast = 60$~TeV and 0.4 passing muons above it.  In that sample, $E_{\rm dep}^{\rm min} = 28$~TeV.  Therefore, for HESE events, $\gamma_\mu \approx 5$, as reported by \Refe\ \cite{Palomares-Ruiz:2015mka}.

For MESE events, we compute $\gamma_\mu$ using the muon rates reports in the 2-year IceCube MESE analysis\ \cite{Aartsen:2014muf}.  Figure 8 in \Refe\ \cite{Aartsen:2014muf} reports 64.78 passing muons below $E_{\rm dep}^\ast = 50$~TeV and 0.44 passing muons above it.  In that sample, $E_{\rm dep}^{\rm min} = 0.33$~TeV.  Therefore, for MESE events, $\gamma \approx 2$.

The probability distribution function for the $i$-th event in a sample to have been generated by atmospheric muons is then
$\mathcal{P}_{i,\mu} = E_{{\rm dep},i}^{-\gamma_\mu} / \int d E_{\rm dep} E_{\rm dep}^{-\gamma_\mu}$, where $E_{{\rm dep},i}$ is the deposited energy of the event.


\section{Details of the statistical analysis}
\label{app:stat_analysis}

\subsection{Prior distributions of the likelihood parameters}
\label{app:likelihood}

Table \ref{tab:priors} shows the prior probability distributions of the likelihood parameters that we have used in our Bayesian statistical analysis.  For the definition of each parameter, see the main text.  Priors are different for the two public IceCube event samples that we use: the 6-year HESE sample\ \cite{Kopper:2015vzf, IC4yrHESEURL, Kopper:2017zzm} and the 2-year MESE sample\ \cite{Aartsen:2014muf, IC2yrMESEURL} restricted to $E_{\rm dep} \geq 20$~TeV. 

For $m$ and $\log_{10} \Gamma$, we use wide uniform priors to avoid introducing unnecessary bias.  For $\log_{10}(B^\prime/{\rm G})$, we choose ranges based on criteria that we explain below.  For $N_{\rm ast}$, $N_{\rm atm}$, $N_\mu$, and $\gamma$ (in the case of HESE) we use priors informed by the IceCube analyses of the HESE and MESE samples; see the footnotes in Table\ \ref{tab:priors} for details.

The ``signal hypothesis'' in Table \ref{tab:priors} refers to the hypothesis where synchrotron-loss features in the diffuse neutrino flux may exist within the energy range of the sample---where they could be detectable---or above its energy range---where they would not affect the sample---but not below its energy range.  By doing this, we prevent the Bayesian parameter scan from finding high values of $B^\prime$ that would induce synchrotron-loss features at artificially low neutrino energies, below the energy range of the sample.   We ensure this by choosing, for each sample, a prior for $\log_{10}(B^\prime/{\rm G})$ such that the energy where muon synchrotron cooling starts to become important, $E_{\nu, \mu}^{\rm synch}$, is larger than the minimum neutrino energy of the sample, for any value of $\Gamma$, and for neutrinos that come from even the highest redshift of $z_{\max} = 4$.  See the main text for the definition of $E_{\nu, \mu}^{\rm synch}$ and Table\ \ref{tab:priors} for details.

The ``null hypothesis'' in Table \ref{tab:priors} refers to the hypothesis where synchrotron-loss features do not affect the event sample, \ie, they may exist only at energies higher than the maximum energy of the sample.  We ensure this by choosing, for each sample, a prior for $\log_{10}(B^\prime/{\rm G})$ such that $E_{\nu, \mu}^{\rm synch}$ is larger than the maximum energy of the sample, for any value of $\Gamma$, and for neutrinos that come from even the highest redshift of $z_{\max} = 4$.  Only the prior for $\log_{10}(B^\prime/{\rm G})$ is different between the signal and null hypotheses.  See Table\ \ref{tab:priors} for details.


\renewcommand{\thetable}{B\arabic{table}}
\setcounter{table}{0}
\squeezetable
\begin{table*}[bt!]
 \begin{ruledtabular}
  \caption{\label{tab:priors}Likelihood parameters varied in our statistical analysis and their prior probability distributions; see \equ{likelihood} in the main text.  See Appendix\ \ref{app:likelihood} for details.}
  \centering
  \begin{tabular}{ccccccc}
   \multirow{2}{*}{Parameter}       & \multicolumn{3}{c}{HESE}  & \multicolumn{3}{c}{MESE ($E_{\rm dep} \geq 20$~TeV)} \\
                                    & Signal hypothesis         & Null hypothesis     & Ref. & Signal hypothesis  & Null hypothesis & Ref. \\ 
   \hline
   $\gamma$                         & Normal $2.92 \pm 0.33$
                                    & Normal $2.92 \pm 0.33$
                                    & \cite{Kopper:2017zzm}
                                    & Uniform $[2,3]$  
                                    & Uniform $[2,3]$               
                                    & -- \\
   $m$                              & Uniform $[-1,4]$       
                                    & Uniform $[-1,4]$ 
                                    & --
                                    & Uniform $[-1,4]$ 
                                    & Uniform $[-1,4]$             
                                    & -- \\
   $\log_{10} \Gamma$               & Uniform $[0,3]$        
                                    & Uniform $[0,3]$  
                                    & --
                                    & Uniform $[0,3]$  
                                    & Uniform $[0,3]$                
                                    & -- \\
   $\log_{10}(B^{\prime}/{\rm G})$  & Uniform $[0,4.34+\log_{10} \Gamma]$\footnotemark[1]
                                    & Uniform $[0,2.29+\log_{10} \Gamma]$\footnotemark[2]
                                    & --
                                    & Uniform $[0,4.29+\log_{10} \Gamma]$\footnotemark[3]
                                    & Uniform $[0,2.29+\log_{10} \Gamma]$\footnotemark[2]
                                    & -- \\
   $N_{\rm ast}$                    & Uniform $[0,N_{\rm obs}=80]$   
                                    & Uniform $[0,N_{\rm obs}=80]$ 
                                    & \cite{Kopper:2017zzm}
                                    & Normal $35.29 \pm 5.94$\footnotemark[4]
                                    & Normal $35.29 \pm 5.94$\footnotemark[4]
                                    & \cite{Aartsen:2014muf} \\
   $N_{\rm atm}$                    & Skew-normal $15.6_{-3.9}^{+11.4}$ 
                                    & Skew-normal $15.6_{-3.9}^{+11.4}$ 
                                    & \cite{Kopper:2017zzm}
                                    & Normal $13.92 \pm 3.73$\footnotemark[4]
                                    & Normal $13.92 \pm 3.73$\footnotemark[4]
                                    & \cite{Aartsen:2014muf} \\
   $N_\mu$                          & Normal $25.2 \pm 7.3$ 
                                    & Normal $25.2 \pm 7.3$
                                    & \cite{Kopper:2017zzm}
                                    & Normal $2.83 \pm 1.68$\footnotemark[4]
                                    & Normal $2.83 \pm 1.68$\footnotemark[4]
                                    & \cite{Aartsen:2014muf}         \\
   \end{tabular}
 \end{ruledtabular}
 \footnotetext[1]{This ensures that synchrotron-loss features, if any, appear in the diffuse flux only at energies $E_{\nu, \mu}^{\rm synch} \geq (1+z_{\max}) E_{\nu, {\rm HESE}}^{\min}$, \ie, not below the energy window of the HESE sample.  We approximate the minimum neutrino energy that could be affected by synchrotron losses by equating it to the smallest HESE deposited energy, \ie, $E_{\nu, {\rm HESE}}^{\min} = 18$~TeV.}
 \footnotetext[2]{This restricts synchrotron-loss features in the diffuse flux to appear only at energies $E_{\nu, \mu}^{\rm synch} \geq (1+z_{\max}) E_{\nu, {\rm HESE}}^{\max} = 10$~PeV, beyond the energy windows of the HESE and MESE sample.  We approximate the maximum neutrino energy that could be affected by synchrotron losses by equating it to the largest HESE deposited energy, \ie, $E_{\nu, {\rm HESE}}^{\max} = 2$~PeV.  (The maximum MESE deposited is 1.3~PeV, but we use the same prior for MESE and for HESE, since the difference is small.) }
 \footnotetext[3]{This ensures that synchrotron-loss features, if any, appear in the diffuse flux only at energies $E_{\nu, \mu}^{\rm synch} \geq (1+z_{\max}) E_{\nu, {\rm MESE}}^{\min}$, \ie, not below the energy window of the MESE sample selected for $E_{\rm dep} \geq 20$~TeV.  We approximate the minimum neutrino energy that could be affected by synchrotron losses by equating it to the smallest MESE deposited energy, \ie, $E_{\nu, {\rm MESE}}^{\min} = 20$~TeV.}
 \footnotetext[4]{For $N_{\rm ast}$, $N_{\rm atm}$, and $N_{\rm mu}$, the central value of each is inferred from interpolating their best-fit contribution to the measured MESE event rates in Fig.\ 8 of \Refe\ \cite{Aartsen:2014muf}.  We count only MESE events with $E_{\rm dep} \geq 20$~TeV.  For each parameter, the standard deviation is assumed to be that for a normal distribution, \ie, the square root of the central value.  For $N_\mu$, we make sure that its sampled values are always non-negative.}
\end{table*}

\subsection{Posterior distributions of the likelihood parameters}

Figures\ \ref{fig:posterior_hese} and \ref{fig:posterior_mese} show the resulting one-dimensional and two-dimensional marginalized posterior probability distributions of the likelihood parameters, obtained under the signal hypothesis, using the HESE and MESE samples.  The posteriors obtained under the null hypothesis (not shown) are similar.  In the main text we discuss the posterior allowed ranges of $\log_{10} \Gamma$ and $\log_{10}(B^\prime/{\rm G})$.   Below, we comment on the ranges of the remaining parameters.

The posterior allowed ranges of $\gamma$, $N_{\rm ast}$, $N_{\rm atm}$, and $N_\mu$ are compatible with the values found in the IceCube analyses of the 6-year HESE sample\ \cite{Kopper:2017zzm} and the 2-year MESE sample\ \cite{Aartsen:2014muf}.  The one instance where this is not the case is for $\gamma$ in the MESE sample: the value that we find, $\gamma \approx 2.12$, is lower than the value reported by IceCube, $\gamma \approx 2.46$.  This is because we only use MESE events with $E_{\rm dep} \geq 20$~TeV, which have a flatter energy distribution than the full sample.

The posterior allowed ranges of $m$ are wide, which reflects that our analysis is only weakly sensitive to this parameter. Our analysis finds a preference for positive values of $m$, \ie, for a number density of sources $\rho$ that grows with redshift up to $z_c = 1.5$ (see the main text), but otherwise does not constrain the value of $m$.  Varying $m$ affects the diffuse neutrino flux in two ways.  First, it changes the flux normalization.  However, because the flux normalization cancels out when computing the partial likelihoods (see the main text), we are not sensitive to this effect.  Second, varying $m$ slightly alters the shape of the spectrum by shifting the energies at which the synchrotron-loss features appear, in the frame of the observer.  However, because the shift is small, our analysis is almost insensitive to the value of $m$.  

As part of our maximization procedure, we compute the evidence---\ie, the likelihood marginalized over all of the parameters---under the signal and null hypotheses, $Z_{\rm signal}$ and $Z_{\rm null}$, respectively.  With them, we compute the Bayes factor $K \equiv Z_{\rm signal}/Z_{\rm null}$ for each sample, to estimate the strength of the evidence in favor of the existence of synchrotron-loss effects in the sample.  We qualify the strength of the evidence using Jeffreys' empirical scale\ \cite{Jeffreys:1939xee}.  For both samples, $\log_{10} K \approx 0.3$, so the strength of the evidence is insignificant, or ``not worth more than a bare mention,'' according to the scale.


\subsection{Computing limits}

To compute the Bayes factor above, we allow all of the likelihood parameters in Table\ \ref{tab:priors} to vary simultaneously.  In contrast, to compute the upper limit on $\log_{10}(B^\prime/{\rm G})$ as a function of $\log_{10} \Gamma$, we fix $\log_{10} \Gamma$ to a given value, vary all of the remaining likelihood parameters, and finally marginalize over all of them except for $\log_{10}(B^\prime/{\rm G})$.  Figure\ \ref{fig:limits} shows the resulting one-dimensional 95\%~C.L.\ upper limit on $\log_{10}(B^\prime/{\rm G})$ as a function of $\log_{10} \Gamma$.

\renewcommand{\thefigure}{A\arabic{figure}}
\setcounter{figure}{0}
\begin{figure*}[t!]
 \centering
 \includegraphics[width=\textwidth]{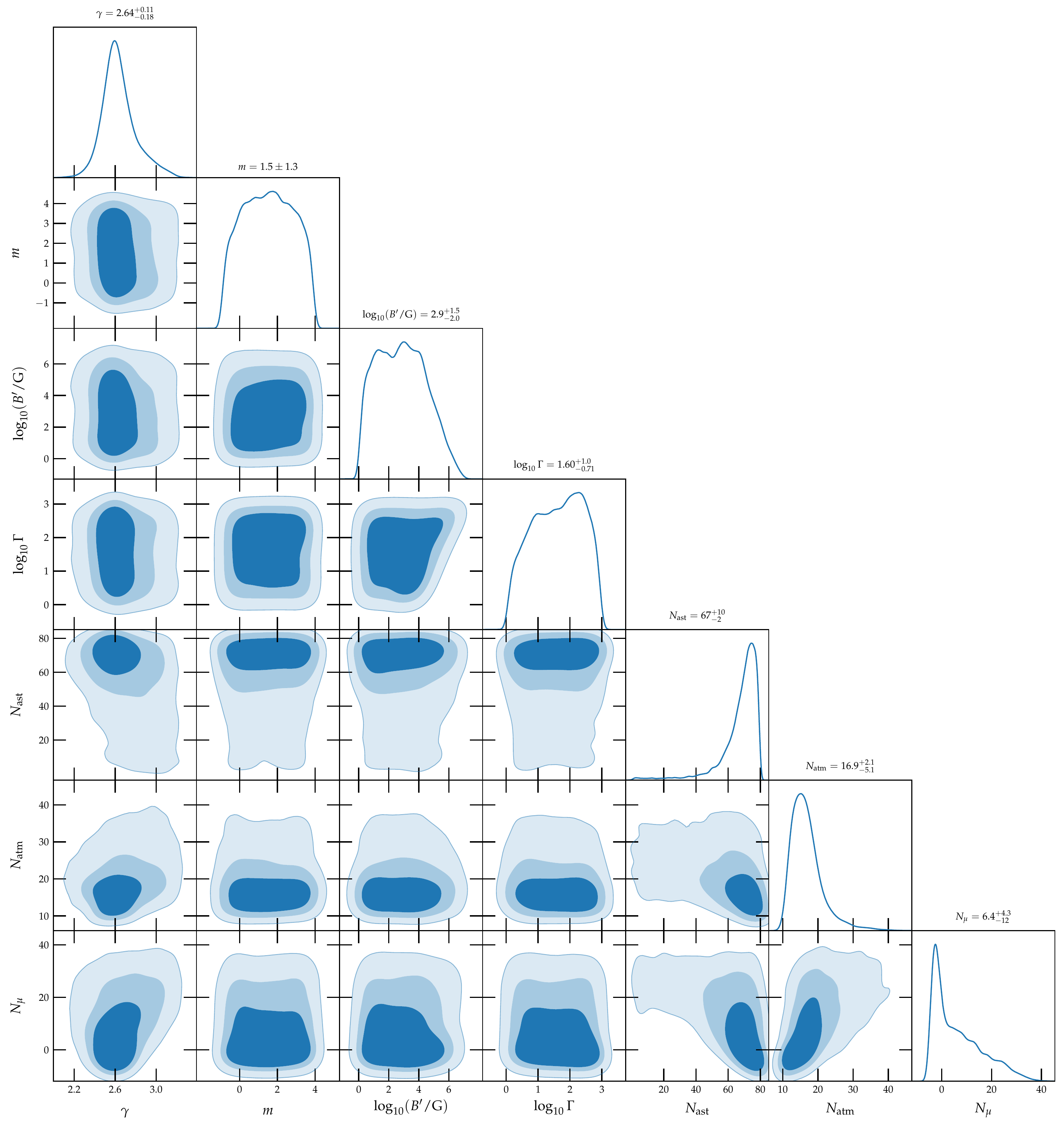}
 \caption{\label{fig:posterior_hese}Posterior probability distributions, central values, and standard deviations of the likelihood parameters for the 6-year IceCube HESE sample, obtained under the signal hypothesis; see Table\ \ref{tab:priors}.  The shaded regions show the 68\%, 90\%, and 95\%~C.L.\ regions, from darkest to lightest shading.}
\end{figure*}

\begin{figure*}[t!]
 \centering
 \includegraphics[width=\textwidth]{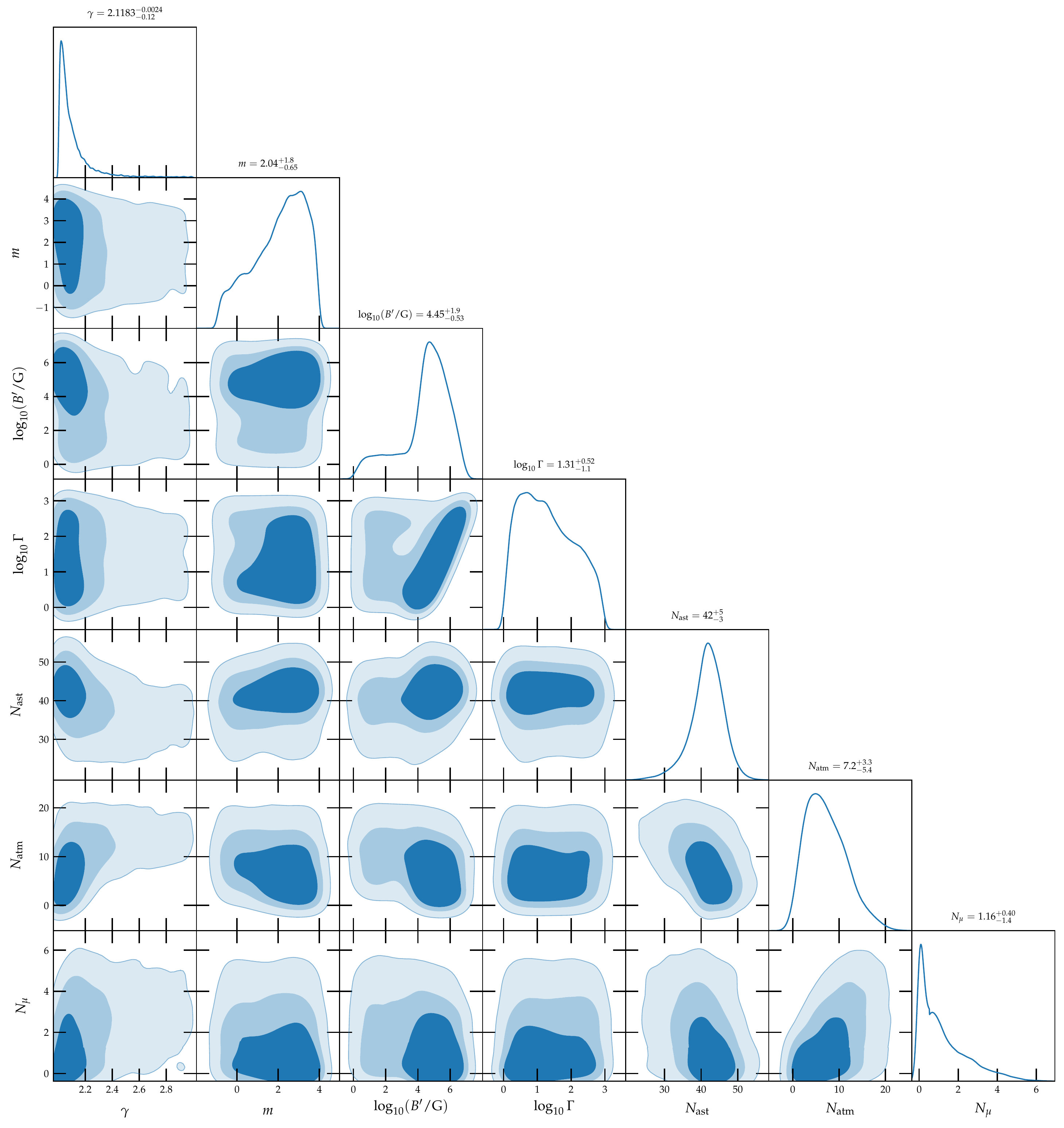}
 \caption{\label{fig:posterior_mese}Same as \figu{posterior_hese}, but for the 2-year IceCube MESE sample.}
\end{figure*}


\clearpage
\newpage

%

\end{document}